\documentclass{article}

\title{Origin of Cosmic Rays}

\author{Luke O'C. Drury\\
\\
{\eiad Institiu/id Ard-Le/inn Bhaile A/tha Cliath}\\
{\eiad Scoil na Fisice Cosmai/, 31 Plas Mhic Liam, Baile A/tha Cliath 2, E/ire}
\\ \\
Dublin Institute for Advanced Studies, School of Cosmic Physics,\\
31 Fitzwilliam Place, Dublin 2, Ireland}

\usepackage{aas_macros}

\usepackage{amsmath}

\usepackage{geometry}                
\usepackage{fullpage}  
\usepackage[parfill]{parskip}  

\usepackage{fourier} % math & rm
\usepackage[scaled=0.875]{helvet} % ss
\newcommand*\eiad{\fontencoding{OT1}\fontfamily{eiad}\selectfont}
\usepackage[OT1,T1]{fontenc}

\usepackage[
colorlinks=true,
linkcolor=blue]{hyperref}

\usepackage[
style=numeric,            % numbered citations
backref=true               % link back to page where cited
]{biblatex}  

\begin{document}

\maketitle

\tableofcontents

\section*{Abstract}

It is argued that there are three `origins' of cosmic rays; the origin of the particles, the origin of the energy, and the site of the acceleration.  The evidence for each origin is discussed and a plausible synthesis outlined for the particles of Galactic origin where the energy comes mainly (but not exclusively) from supernova explosions, the site of the acceleration is at strong collisionless shock waves, and the accelerated particles come from the interstellar and circumstellar material swept over by these shocks.  If these shocks are capable (as indicated by recent observations and theoretical work) of significantly amplifying magnetic fields this picture appears capable of explaining the cosmic ray particles at all energies below the `ankle' at $3\times10^{18}\,\rm eV$.  The particles above this energy are generally taken to be of extra-galactic origin and possible acceleration sites for these UHE particles are briefly discussed.

\twocolumn
\sloppy

\section{Introduction}

Review articles with this, or very similar, titles have been published for nearly a century now and the phrase is familiar as the title of the influential monograph by Ginzburg and Syrovatskii \cite{Ginzburg:1969lr} (which is still well worth reading). Indeed a search of the ADS database for titles including these words returned 1009 hits!  However despite its deceptive simplicity we still do not have a definitive and universally accepted answer to the implicit question it poses.  In part this is because there is a level of ambiguity in the word "origin" and there are at least three distinct meanings that one could attribute to it, and thus three different questions that are being asked.  Observationally it is well established that the cosmic rays are ordinary atomic nuclei accelerated to very high energies which arrive in the neighbourhood of the Earth from outside the solar system, together with some electrons and secondary particles.  In asking for the origin of these particles we need to distinguish between the source of the matter from which these nuclei come, the source of the energy which powers their acceleration, and the location of the physical system in which the acceleration occurs.  Logically these are three distinct questions with potentially three different answers (indeed I will argue that this is the case).  Some confusion has been caused in the past by a lack of clarity on this point.   Of course as a scientist one would hope to have a model which answered all three questions  in a manner which was also fully consistent with our general understanding of astronomy and knowledge of the cosmos.  We are perhaps very close to this goal, but we are certainly not quite there yet.

\section{Galactic origins?}

One thing on which there is universal agreement is that the bulk of the cosmic rays observed at mildly relativistic energies, around $1\,\rm GeV$ per nucleon, are of exclusively Galactic origin. Apart from the inherent implausibility and energetic difficulty of supposing a sea of cosmic rays to permeate the entire universe, at these energies the distribution of cosmic rays in the Galaxy and in its dwarf satellites can be determined rather directly by gamma-ray observations.  The key point is that the diffuse gamma-ray emission of the Galaxy above about $100\,\rm MeV$ is dominated by emission from the decay of neutral pions.  These in turn are produced in hadronic collisions between cosmic ray nuclei and nuclei in the interstellar medium.  Thus if one knows the gas distribution and observes the diffuse gamma-ray emission (as determined by a series of ever better satellite measurements starting with COS-B \cite{1986A&A...154...25B}, CGRO \cite{1996A&A...308L..21S}, AGILE \cite{2004MSAIS...5..135G} and most recently FERMI \cite{2009PhRvL.103y1101A}) it is possible to infer the spatial distribution of cosmic rays at around $1\,\rm GeV$ per nucleon.  The results show clearly that these mildly relativistic cosmic rays are rather smoothly distributed throughout the disc of the Galaxy, but that there is a definite radial gradient and that the intensity is higher in the inner Galaxy and falls off towards the outer disc.  This alone would be conclusive proof that the particles were being produced in the Galaxy rather than diffusing in from outside (it should  be said for completeness that although there is a clear radial gradient it is actually less than one would naively expect; this may be explicable by dynamical feed-back effects \cite{2002A&A...385..216B}). Perhaps the most decisive result though is that the level of cosmic rays in the Magellanic clouds is definitely substantially lower than that in the Galaxy \cite{1993PhRvL..70..127S, 2010arXiv1008.2127T} which clearly rules out the idea that the Galaxy and its satellites are bathed in a universal sea of cosmic ray particles.

Of course the fact that the mildly relativistic particles are of Galactic origin does not automatically mean that all the particles at higher energies are also Galactic, and indeed there are good theoretical and observational reasons to think that at the very highest energies we are in fact seeing an extra-galactic component.  However the bulk of the cosmic rays, certainly up to the so-called ``knee'' at $3\times10^{15}\,\rm eV$ and probably as far as the ``ankle'' at $3\times 10^{18}\,\rm eV$ are thought to be of local Galactic origin.  To suppose that the particles between the ``knee'' and the ``ankle'' are not part of the same Galactic population requires artificial fine-tuning to make the two populations join smoothly at the ``knee'' in such a way that the spectrum steepens there; this should be contrasted with the hardening at the ``ankle'' which can easily be fit as a harder component coming in.
The arguments for an extra-galactic origin at the very highest energy are firstly, that at these energies the gyroradii of the particles in a standard Galactic magnetic field of $0.3\,\rm nT$ are comparable to the thickness of the Galaxy (at the ``ankle'' a proton would have a gyroradius of $1\,\rm kpc$) so that even if they could be produced in the Galaxy they would not be confined to it, secondly that the hardening of the spectrum in this region is suggestive of a new component, and finally that there is evidence of a correlation of arrival directions at the highest energies with the large-scale distribution of matter in the near-by universe.  

\section{Origin of the accelerated particles}

Let us start by considering the question of where the particles themselves come from, the first interpretation listed above.  Out of what reservoir of material are the atomic nuclei drawn which ultimately end up as the cosmic rays that we observe?  The standard approach to such a question is to look in detail at the chemical and isotopic composition in the hope of finding a clear indication of some specific source.  Indeed in the past much effort was expended in attempting to measure such exotic aspects as the ultra-heavy nuclear abundances in the cosmic rays in the hope that this would link them to recent r-process nucleosynthesis in supernovae.  In fact there is some evidence for modest enhancements of the actinides and other ultra-heavy elements (Donnelly et al, submitted) but there is no clear link to a specific and unique composition which could tie down the origin unambiguously.  On the contrary the composition (see below for a discussion of what this means) is in broad terms disappointingly normal and not too different from the ``standard solar-system abundances'' which are also taken to be representative of standard Galactic abundances in the local ISM.

More precisely, by the chemical composition of the Galactic cosmic rays we usually mean the relative abundances of the various nuclear species measured at the same energy per nucleon.  From a theoretical point of view it might be better to measure at the same rigidity (momentum per charge) but the difference is unimportant for all nuclei apart from hydrogen because to a good approximation, apart from hydrogen, they all have roughly equal numbers of protons and neutrons.  Such measurements are technically quite easy for the mildly relativistic particles with energy per nucleon of a few GeV where the fluxes are also sufficient to allow good statistics with experiments of modest scale.  It is very much harder to push the measurements to higher energies, partially because of the rapidly falling fluxes, but also because particle identification is technically more challenging in the relativistic regime.  However there has been recent progress here with interesting results in particular from the PAMELA collaboration.  It is also difficult to go to lower energies because the heliospheric effects of solar modulation, solar energetic particles and anomalous cosmic rays overwhelm the signal of the Galactic cosmic rays.  However for mildly relativistic cosmic rays (and it is worth remembering that this is the energetically dominant component) we now have a large body of evidence accumulated by numerous experiments over many years.

The first point to make is that in talking about the composition we are implicitly assuming that all the species have identical energy spectra; otherwise the composition would be energy dependent and it would be necessary to specify what reference energy per nucleon one was using.  In fact to a good approximation this appears to be the case, although there are very interesting recent measurements which suggest that there are slight, but significant, differences between in particular the hydrogen and helium energy spectra and that the composition thus has a slight energy dependence.  Ignoring this, and correcting for the well-known effects of interstellar propagation, the inferred source composition (that is the composition of the beam coming out of the cosmic ray accelerator) is rather normal.  All the nuclei up to Uranium (and one plausible trans-uranic actinide candidate) have been seen in the cosmic rays in proportions that are generally close to what one would get from a well-mixed sample of the local Galaxy (for which the primordial solar system composition is usually taken as a proxy).
However superimposed on this there are some significant effects.  In particular Iron and most other heavy elements are clearly over-abundant relative to the light elements such as hydrogen and helium by about a factor of 30.  Because these enhancements appear to correlate with first ionization potential (FIP), and because a similar effect is know to operate in the solar energetic particles, this effect was long interpreted as a FIP-based bias.  However no satisfactory physical model for the supposed FIP-effect exists (for cosmic rays) and thus the result should be seen purely as an empirical correlation \cite{1997SSRv...81..107W}.  In fact a careful examination of the pattern of over-abundances suggests that no single-parameter model can explain the observations and that a two-parameter model at least is required \cite{Meyer:1997lr}.    The fact that the overabundant elements are mainly refractory and are expected to be locked up in dust grains throughout most of the interstellar medium at first sight just adds to the mystery, but may in fact be the key to understanding the data as suggested in \cite{Ellison:1997fk} where it is shown that a contribution from sputtering of accelerated dust operating in parallel with a direct gas-phase injection process can rather naturally explain all features of the composition.

It is perhaps worth pointing out that the relative normality of the composition already rules out some exotic models for cosmic ray origin.  It clearly does not resemble, for example, the ground-up iron and nickel composition one would expect from a neutron star crust so that thermionic emission from pulsars can be ruled out as the dominant source (not that this was ever a very plausible model anyway).  Similarly direct production in one particular sub-class of supernova is ruled out; the composition, like the general composition of the Galaxy, requires the mixing of a variety of nucleosynthetic components \cite{1995AdSpR..15Q..35S} produced in different nucleosynthetic sites.

The one clear hint of a non-standard composition is the well-documented excess of $^{22}$Ne in the cosmic rays.  The isotopic ratio $^{22}$Ne/$^{20}$Ne measured in the GCR has a value about 5 times that in the solar wind  \cite{2008NewAR..52..427B} and this is the only clear isotopic (as distinct from compositional) anomaly observed in the cosmic rays.  This is suggestive of a link to winds from Wolf-Rayet stars, thought to be the major source of $^{22}$Ne, and certainly suggests a link to OB associations and/or super-bubbles.

It is worth pointing out that the electron component in the cosmic rays clearly has a very different spectrum and propagation history to the nuclear components and thus statements about the proton to electron ratio need to be heavily qualified.  It is not obvious that comparing fluxes of GeV electrons with protons at the same kinetic energy per particle is a particularly sensible thing to do, but on this measure the electrons are significantly underabundant in the cosmic rays with a flux only a few percent that of protons.  Combined with the tendency for an overabundance of heavy elements this suggests that the cosmic ray accelerator preferentially accelerates `heavy' particles although this should be treated with some caution.    A final point worth making about the electrons is that at TeV energies the radiative losses are such that these particles must come from relatively near-by sources
\cite{1997AdSpR..19..767N,1995A&A...294L..41A}.

The final point to be made is that the detailed pattern of deviations from ``standard'' abundances in the arriving cosmic rays must be explained by any account of cosmic ray origin.  There is substantial information-content in this data, even if its interpretation is not easy, and it significantly constrains models of cosmic ray origin.  The recent very interesting observations of small deviations from the first-order model of a single power-law for all elements suggest that the data are now of sufficient quality to allow higher-order corrections to the basic theory to be measured with interesting implications for theory, see e.g.~\cite{2011MNRAS.415.1807D,2011ApJ...729L..13O}

\section{Origin of the energy}

By contrast to the lack of clarity about the composition the situation as regards the origin of the energy is simpler.  Standard ideas on cosmic ray propagation in the Galaxy (based largely on constraints from secondary production in spallation reactions as well as life-time estimates derived from radioactive secondaries) lead to the unambiguous conclusion that the power needed to maintain the cosmic ray population of the Galaxy must be about  $10^{41} \,\rm erg\,s^{-1}= 10^{34}\,W$;  indeed a recent analysis \cite{Strong:2010lr} using the Galprop model for cosmic ray propagation estimates that the total cosmic ray input luminosity for the Galaxy is $(0.7\pm0.1)\times 10^{34}\,\rm W$ where the uncertainty reflects the allowed range of propagation model parameters within the Galprop framework (the true uncertainty must thus be larger; it is remarkable how tightly constrained the Galprop luminosity is). 
It is interesting to note that Ginzburg and Syrovatskii even as far back as 1964 \cite{Ginzburg:1969lr} (p 191) conservatively estimated that ``In a quasi-stationary state the power of the sources supplying cosmic rays to the Galaxy must be $U_{\rm Source}\approx 3\times 10^{40}\,\rm erg\,s^{-1}$'' which is only a factor of two less than the Galprop result.  As noted in the appendix to \cite{1989A&A...225..179D} the conservative estimate of $3\times 10^{33}\,\rm W$ ignores the energy dependence of the escape time, and for plausible power-law dependencies this increases the power requirement by a factor between 3.5 and 10 giving estimates of between $10^{34}\,\rm W$ and $3\times 10^{34}\,\rm W$ (note that the upper estimate is based on harder source spectra and stronger energy dependent escape than
allowed in Galprop).  It seems certain that the Galactic cosmic ray luminosity is thus at least $6\times 10^{33}\,\rm W$ and at most $3\times 10^{34}\,\rm W$.

The question then is what energy sources in the Galaxy are powerful enough to run an accelerator producing this output beam power?  The standard answer, which has been given for the last half century, is that the only plausible energy source is the explosion of supernovae.  Indeed if the mechanical energy released per supernova is of order $10^{44}\,\rm J$ and they occur at a rate of one every 30 years or so, the power available in bulk motion is some $10^{35}\,\rm W$.  Thus as long as an acceleration process exists which can convert about 10\% of this energy into accelerated particles supernovae are a viable power source.
It is worth noting that the automated supernova searches currently being undertaken by cosmologists to study dark energy will give us a greatly improved knowledge of supernova statistics and properties.  The recently discovered ultra-luminous supernovae \cite{2011Natur.474..487Q, 2011arXiv1107.3552C} as well as anomalously powerful supernovae \cite{2011ApJ...728...14P} give some indication of what may be discovered.

In fact supernovae are thought to be almost the only available power source.  The key point is that the energy has to be in a form that is capable of driving particle acceleration and this means that it must be either magnetic field energy (driving acceleration through magnetic reconnection) or kinetic energy (driving Fermi acceleration).  The bulk radiative luminosity of all the stars in the Galaxy, for example, is substantially larger at about $10^{37}\,\rm W$ \cite{springerlink:10.1007/s001590050019}, but there is no known way for an astrophysical particle acceleration process to tap into this general photon field.  Similarly the bulk of the energy released in core-collapse supernovae is thought to escape in the form of neutrinos and gravitational waves with only of order 1\% going into the mechanical explosion, but there is no known way to use this neutrino and gravitational wave energy to drive an accelerator.

Pulsars come a close second to supernovae, but are about an order of magnitude less powerful.  It is in fact very plausible that pulsar driven magnetic winds may account for some of the electrons and positrons seen in the cosmic rays at high energies as suggested by recent data \cite{2011APh....34..528D} (in particular the Pamela positron excess), but they cannot account for the bulk of the atomic nuclei on energetic grounds alone (quite apart from the compositional issues discussed in the previous section).  Similar remarks apply to the strong winds from OB stars.  These may contribute at the level of a few percent, but are not thought to be strong enough to produce the bulk of the observed cosmic rays.   It is interesting to observe that the well-known solar modulation of low-energy cosmic rays shows that the solar wind, in pushing these particles out of the inner heliosphere, must in fact be doing work on them and thereby contributing in a small way to the acceleration at low energies, but the amount of energy involved is trivial compared to the energy budget needed, even when integrated over all low-mass stars in the Galaxy.  The power of the Solar wind is of order $3\times 10^{20}\,\rm W$ so that even for $10^{11}$ solar-type stars in the Galaxy the total possible contribution is only $3\times10^{31}\,\rm W$ or less than 1\% of the cosmic ray luminosity.

A further complication, ofter overlooked, is that if the energy were to be transferred promptly to accelerated particles soon after the supernova explosion, the subsequent adiabatic losses of the particles as they drive an expanding bubble into the surrounding medium would place an impossible energy demand on the accelerator.  It follows that while supernova explosions may be the ultimate source of the energy, the acceleration process itself must take place at later times and in the supernova remnant that forms around the explosion site.  Another argument for this is to observe that energy can only be efficiently extracted from the bulk motion of the ejecta when the ejecta collide with roughly the same amount of stationary material; it is the differential motion between roughly equal masses that is the source of free energy to drive the accelerator and thus it is only after the ejecta have swept up roughly the same mass of circumstellar material that peak power is available.  In the case of Galactic supernovae this typically takes a few hundred years.  Particle acceleration at earlier times is certainly possible (and indeed is required by radio observations, eg \cite{2007AIPC..924..407P}) but the total power available is limited. 
For a recent calculation including a careful treatment of the adiabatic losses see
\cite{2010APh....33..160C}.

\section{The acceleration site and mechanism}

The final implicit question posed by the title of this article is that of how and where the particles are accelerated and here there has been very significant progress in the last few decades.  The key development, starting with four seminal articles in 1977 and 1978 \cite{1977DoSSR.234Q1306K, 1977ICRC...11..132A,1978MNRAS.182..147B, 1978ApJ...221L..29B}, was the realisation that a version of Fermi acceleration operating at shock waves could naturally produce power-law spectra of accelerated particles with the characteristics required to explain the origin of the Galactic cosmic rays.  This process, now generally referred to as diffusive shock acceleration (DSA), is an inevitable consequence of assuming that the propagation of energetic charged particles can be modelled as essentially a random walk (ie a diffusion process) in which particles conserve their energy in the local fluid rest frame together with the idea that the shock front can be represented as an abrupt compressive discontinuity in the velocity of this rest frame.  As perhaps most clearly elucidated by Bell these two ideas require that particles gain energy each time they diffuse across the shock, and that if the shock is non-relativistic the particles make very many shock crossings before being finally advected downstream.  The end result is that a power-law spectrum (in momentum) of accelerated particles is formed stretching from whatever initial starting energy the particles have up to a maximum energy determined by the finite age and size of the shock.  There are a number of review articles summarising the theory of diffusive shock acceleration which can be consulted for further details \cite{1983RPPh...46..973D,2001RPPh...64..429M, 1994ApJS...90..561J, 1987PhR...154....1B, 1991SSRv...58..259J}.

This process has a number of key features which explain why it is now central to most discussions of cosmic ray origin.  Firstly, it is very natural and depends only on rather robust and simple physics; that particles are scattered in direction, but not in energy, by magnetic fields; that the magnetic fields are tied to the plasma; and that the plasma is discontinuously compressed in shock fronts.  

Secondly, it produces power-law spectra without any unnatural fine-tuning (unlike most other variants of Fermi acceleration); the exponent of the power-law is fixed entirely by the compression ratio of the shock and for standard shock compressions of four the simple theory predicts a universal energy spectrum at relativistic energies of the form $N(E)\propto E^{-2}$, close to what is inferred for the cosmic ray source spectrum.  Of course if the shock puts a significant amount of the available energy into cosmic rays the simple linear theory is not applicable.  The nonlinear theory predicts slightly concave spectra which are softer at low energies and harden at high energies, but which share the $E^{-2}$ feature of having comparable amounts of energy in the low and high energy components.

Thirdly, and again unlike other versions of Fermi acceleration, it does not require a separate pre-acceleration phase to produce seed particles for further acceleration; the process appears capable of accelerating particles directly from the thermal population all the way up to the highest energies allowed by the scale of the shock.  

Fourthly, there appears no reason why the process could not operate at high efficiencies; of course in this limit reaction effects have to be considered and the simple linear theory clearly breaks down, but it is plausible that cosmic ray acceleration provides a major part of the energy dissipation in strong collisionless shocks under astrophysical conditions \cite{1993ARA&A..31..373D} (essentially because the large scattering mean free path of the cosmic ray particles makes them the most effective agents in providing the necessary dissipation).

It is perhaps worth pointing out that at the micro-level all acceleration processes, as indeed all physical laws, are time-reversible and thus as long as processes are adiabatic, any energy gain can be turned into an exactly compensating energy loss.  To achieve real acceleration it is essential to have an entropy increasing process which is irreversible in character.  In the case of shock acceleration this stochastic element comes from the random scattering of the particles leading to a diffusive spatial transport.  It is also significant that shocks are essentially self-forming dissipative structures where nature transfers energy from ordered bulk motion to random motion of individual particles and high-frequency wave modes. In conventional shock theory it is assumed that the energy ends up as thermal energy, but as pointed out long ago by Hoyle \cite{1960MNRAS.120..338H}, under astrophysical conditions there are three possible energy reservoirs; thermal particles, magnetic fields and non-thermal particles (or cosmic rays).  Indeed as has often been commented on, in the interstellar medium all three appear to have roughly equal values.  Hoyle presciently observed that there seemed to be no reason for a collisionless shock under astrophysical conditions to favour any one reservoir as the preferred sink for the kinetic energy being dissipated in the shock.  This is in essence our current understanding.  Collisionless shocks under astrophysical conditions are thought to transfer energy efficiently into plasma heating, into accelerated particles and into an amplification and tangling of the magnetic field.  The big advance from Hoyle is that we now have identified explicit physical mechanisms for some of these energy transfers and at least in principle can begin to calculate the relative amounts of energy transferred into each channel as a function of shock parameters (although this is still very much work in progress).

%Discuss GBR energetics?

\section{Physical limits on the accelerator}

There are a number of important constraints on the acceleration process which act to limit the maximum attainable energy.  The first and simplest is that in any accelerator where the particles are magnetically confined while being accelerated the gyroradius of the particles has to be less that the size of the system.  Thus for relativistic particles of momentum $p$ and energy $E=cp$, in any accelerator of size $R$ with magnetic fields of strength $B$ we have,
\begin{equation}
r_g = {p\over eB} = {E\over ecB} < R \implies E<ecBR.
\end{equation}
This is of course an upper bound and the actual maximum attainable energy will generally be lower.

In the case of diffusive shock acceleration the theory requires, among other things, that the diffusion length-scale of the particles be small compared to the shock radius,
\begin{equation}
{\kappa(p)\over \dot R} \ll R
\end{equation}
where $\kappa$ is the diffusion coefficient,  $\dot R$ is the shock speed and $R$ is the shock radius (this is for the case of spherical shock expanding into a stationary medium).  If we make the usual assumption that the lowest value the diffusion coefficient can have is the Bohm limit (mean free path comparable to the gyroradius) then this gives the limit,
\begin{equation}
{1\over3} r_g c < \kappa \implies r_g \ll {3 R \dot R\over c}
\end{equation}
and thus (dropping the factor 3 for simplicity) the tighter limit
\begin{equation}
E< e B R \dot R.
\end{equation}
This is often, introducing the dimensionless velocity $\beta = \dot R/c$, written in the form
\begin{equation}
E < e c \beta R B
\end{equation}
and referred to as the Hillas limit in reference to the well-know Hillas plot where various astrophysical systems are plotted on a $B,R$ plane \cite{1984ARA&A..22..425H}.

In addition to these limits from the finite size, there are limitations from the finite age of the system; the accelerator has to run long enough to get particles from the injection energy up to the required energy.  A well know result of linear shock acceleration theory is that the acceleration time scale is given by
\begin{equation}
t_{\rm acc} = {3\over U_1 - U_2} \left({\kappa_1\over U_1} + {\kappa_2\over U_2}\right)
\end{equation}
where $U_1, U_2$ are the upstream and downstream flow velocities and $\kappa_1, \kappa_2$ the corresponding diffusion coefficients (assumed spatially constant).  The extension to the case of arbitrary spatial dependence of the diffusion coefficient is relatively straightforward \cite{1991MNRAS.251..340D}.
To order of magnitude this shows that
\begin{equation}
t_{\rm acc} \approx 10 {\kappa\over \dot R^2}
\end{equation}
and thus if the acceleration time scale is to be less than the dynamical time of the system,
\begin{equation}
t_{\rm acc} < {R\over \dot R} \implies \kappa < {1\over 10} R \dot R
\end{equation}
If the diffusion coefficient has Bohm scaling (scattering mean free path comparable to the gyro radius) then within factors of order unity this is thus the same as the Hillas limit although somewhat stricter when the numerical factors are included.  Of course the acceleration time-scale must also be short compared to that of any competing energy-loss process, but this is normally only of concern for electrons (although on cosmological scales it may also be important for protons in galaxy clusters
\cite{2010PhyU...53..691P}).

The important point to make is that diffusive shock acceleration provides a concrete acceleration process which saturates the Hillas limit if the diffusion can be driven down to values of order the Bohm limit.  Thus if the magnetic field is sufficiently tangled on the relevant scales for the scattering mean free path of charged particles to be comparable to the gyroradius, then the maximum rigidity to which particles are accelerated is of order the length scale of the system times the velocity scale times the effective magnetic field strength.
If we take fairly standard values for a SNR shock; a radius of a few parsecs, velocity of 1\% the speed of light and a standard ISM magnetic field of $0.3\,\rm nT$ this gives a maximum particle rigidity of about $10^{14}\,\rm V$.  As pointed out long ago \cite{1983A&A...125..249L} this is tantalisingly close to, but definitely a bit short, of the ``knee'' at $3\times 10^{15}\,\rm V$.  Of course ideally we want acceleration in the Galaxy to operate as far as the `ankle' and not just stop at the `knee' so this is a real problem.  The problem can be somewhat alleviated by noting that the `knee' and the `ankle' are features measured in the all-particle energy spectrum and that if the composition is quite heavy with a substantial contribution from Iron nuclei (for which there is some evidence) this gains one a factor of 26 relative to a pure proton composition.  In fact it is quite plausible that some part of the decline from the `knee' to the `ankle' is just a reflection of the chemical composition, but one still need an acceleration process that can accelerate to a particle rigidity of $6\times 10^{16}\,\rm V$ if the Iron energy spectrum is to extend to $10^{18}\,\rm eV$ \cite{Aloisio:2006lr} although it is possible to fit the data with models where the Galactic component cuts off at rigidities as low as $6\times 10^{15}\,\rm eV$ \cite{Blasi:2012qy}.

It is interesting to note that for a Sedov-Taylor self-similar blast wave with $R\propto t^{2/5}$ and thus $\dot R \propto T^{-3/5}$ the product $R\dot R \propto t^{-1/5}$ is almost constant with only a very weak decrease with time.  If we evaluate the product at sweep-up, then we have
\begin{equation}
\rho R^3 \approx M_{\rm ej},\qquad E_{\rm SN} \approx M_{\rm ej} \dot R^2 \implies
R\dot R \approx \left(E_{\rm  SN}\over\rho\right)^{1/2} M_{\rm ej}^{-1/6}
\end{equation}
and the product is also very weakly dependent on the ejecta mass $M_{\rm ej}$ and has only a square root dependence on the explosion energy $E_{\rm SN}$ and ambient density $\rho$.
It is thus essentially impossible to gain the several orders of magnitude needed if we want to push the Hillas limit up to the `ankle' by manipulating $R\dot R$ and the only hope is to increase the magnetic field $B$.

The big break through in the last decade or so has been the realisation, starting with the seminal work of Bell, \cite{2004MNRAS.353..550B} and references therein, that the effective field strength at the shock may be substantially larger than the standard ISM field, and that in this way it may be possible to push acceleration in Galactic shocks up to substantially higher energies and possibly even as far as the ``ankle''.  In fact there is quite a substantial body of observational evidence pointing to amplified fields in young supernova remnants (e.g. \cite{2008AIPC.1085..169V})
and a number of plausible processes (in addition to that identified by Bell) which can twist and amplify the fields in the shock precursor region, e.g.
\cite{2009ApJ...692.1571M,2010ApJ...717.1054R,2011arXiv1110.0257M}.  Clearly the maximum field amplification possible would wind the field up to close to energy equipartition with the kinetic energy of the flow,
\begin{equation}
{B^2\over 2\mu_0} \approx {1\over 2} \rho U^2.
\end{equation}
More generally we can write
\begin{equation}
{B^2\over 2\mu_0} \approx {1\over 2} \rho U^2 \left(U\over c\right)^\alpha
\end{equation}
where Bell, for example, suggests that his instability saturates with $\alpha=1$.  

If we use this amplified field in the Hillas limit we get, inserting numbers,
\begin{equation}
E< 5\times 10^{21} \beta^{2+\alpha/2} \left(n\over 1\,\rm cm^{-3}\right)^{1/2} \left(R\over 1\,\rm pc\right)\,\rm V
\end{equation}
as the absolute maximum rigidity to which a shock of dimensionless velocity $\beta$ propagating in a medium of hydrogen number density $n$ and length-scale $R$ can accelerate ions in a shock-amplified field.  Of course this is very much an upper limit.  In reality the field is unlikely to be amplified to full equipartition, and even if it does this will be mainly downstream and not upstream (and it is the upstream field that in many ways is critical for the acceleration, \cite{2008ApJ...673L..47E}).  However with this caveat this gives rise to an interesting variant of the classic Hillas plot where instead of plotting objects in the $B,R$ plane they are plotted on the $\rho,R$ plane.  It certainly appears possible for shocks with $\beta\approx 10^{-2}$ ($3000\,\rm km\,s^{-1}$) to accelerate ions to a rigidity in the region of $10^{17}\,\rm V$ in Galactic scale sources \cite{2008ApJ...678..939Z}.  

It should be pointed out that one serious issue often overlooked is that it is not enough to just amplify the field on small length-scales.  While this can lead to a reduced diffusion coefficient for low-energy particles, if the highest energy particles are to be affected the field has to be amplified on scales at least as large as their gyro-radius which, at least for some of the proposed mechanisms, requires an efficient non-linear inverse cascade to transfer energy from the small scales to the large scales.

The other major limitation on the acceleration process is the total amount of energy that can be used to accelerate non-thermal particles.  This is clearly limited by the total mechanical power available in the shock, that is an energy flux per unit surface area of
\begin{equation}
{1\over 2} \rho_1 U_1^3  - {1\over 2} \rho_2 U_2^3 = {1\over 2}\rho_1 U_1^3\left[1-\left(U_2\over U_1\right)^2\right]
\end{equation}
One of the big challenges for shock acceleration theory has been to understand how the process works when a significant amount of this energy goes into the acceleration.  Clearly it is then inconsistent to treat the particles as test particles and their reaction on the system has to be self-consistently included in the dynamics (as indeed was pointed out in the earliest papers on diffusive shock acceleration).

Formulating the problem is easy (in essence one just adds the accelerated particle pressure as an extra term in the momentum equation with corresponding terms in the energy equation) but understanding the behaviour of this coupled system is another matter.  Very substantial progress has been made, in particular through the development of semi-analytic approximations, and there is a feeling that the problem is effectively solved.  However a word of caution is in order. The problem is that all the various approaches which have been used assume that the shock structure can be treated as quasi-stationary and steady on intermediate length scales (that is between the micro-scales of the plasma physics and the macro-scales of the astrophysical system) whereas in reality there is a whole zoo of instabilities operating on these scales.  This is of course good news from the point of view of magnetic field amplification, because it is precisely these instabilities that lead to the growth of the effective scattering field and its entanglement.   The good agreement of the various different approaches to the modified shock structure may thus be somewhat illusory.

With this caveat, what can be said is that it appears entirely possible for diffusive shock acceleration to operate at relatively high efficiency with half or more of the available energy going into accelerated particles, and indeed this appears to be the natural state if the process is self-regulated by the reaction of the accelerated particles.  A down-side of this is that as a consequence of the reaction effects the spectrum is no longer a simple power-law but becomes concave, with a steeper slope at low energies and hardening towards high energies before cutting off.  The effect of particles escaping at the highest energies also has to be allowed for, and indeed in the extreme limit the shock becomes almost a mono-energetic source of particles escaping at the upper cut-off energy.  However as long as the effects are not too great, and as long as the spectrum stays reasonably close to the canonical $N(E)\propto E^{-2}$ it is possible to show that these effects largely average out when integrated over time \cite{Caprioli2010160, 2011MNRAS.415.1807D,2008ApJ...678..939Z}.    It should be pointed out however that such hard production spectra do not fit easily into propagation models constrained by the known observational data.  In particular the low anisotropy is very hard to fit  \cite{Blasi:2012fk,Ptuskin:2006uq} and generally people working on propagation models prefer significantly softer starting spectra, more like $E^{-2.3}$.   It is worth noting that the anisotropy data is rapidly improving in quality with new observations and analyses from {\it inter alia} MILAGRO \cite{Battaner:2009yq}, ARGO \cite{Aielli:2012kx} and ICECUBE \cite{Abbasi:2012fj} becoming available ( a classic example of one person's background being another person's signal).

Finally, a brief word on electrons.  Diffusive shock acceleration is in principle as capable of accelerating electrons as protons, but there are two important differences.  The first, which is relatively trivial, is that the electrons are subject to strong radiative losses which can limit the maximum energy attainable.  The more fundamental difference is that the gap between thermal energies and the energies at which the particles can be considered sufficiently energetic for the approximations of diffusive shock acceleration to apply is very small for protons, but large for electrons.  This has important consequences for the injection process whereby some small fraction of the thermal particles become non-thermal and enter the acceleration process.  It is certain that the injection physics for electrons must be quite different to that for protons and heavy ions.  A very interesting recent suggestion is that the electrons may in fact get injected by, so to speak,  hitching a ride on partially ionized high-charge nuclei \cite{2011MNRAS.412.2333M}.  It turns out that the electron stripping time-scales for heavy ions can be comparable to the acceleration time-scales, so that by accelerating partially-ionized heavy ions and then stripping off the energetic electrons a small population of pre-accelerated electrons can be made available for shock acceleration.  This is reminiscent of the idea that the refractory elements are pre-accelerated in charged dust grains thereby by-passing the strong mass-dependent fractionation seen in gas phase nuclei.  
It is also plausible that pulsars contribution an additional source of equal numbers of positrons and electrons with a relatively hard spectrum which may be responsible for the observed upturn in the positron fraction at high-energies observed by PAMELA \cite{2010APh....34....1A} (although other explanations are certainly possible, eg secondary production within SNRs \cite{2009PhRvL.103e1104B} or production by beta decay of radioactive nuclei, \cite{1993ApJ...405..614C,2011PhRvD..84h3010Z}.

\section{A possible synthesis?}

A plausible account of the origin of Galactic cosmic rays is thus the following.  The accelerated nuclei are the non-thermal tails of the particle distribution functions behind strong collisionless shocks, mostly those bounding supernova remnants.  These distribution functions extend up to a limit given by the Hillas criterion, but with a shock-amplified magnetic field, and have roughly equal amounts of energy per logarithmic interval as well as a total energy content comparable to the kinetic energy density in the inflowing plasma.  At early times when the shocks are at their fastest the maximum particle rigidity may extend up to $10^{17}\,\rm V$ (with the total energy spectrum extending to $3\times 10^{18}\,\rm eV$ when allowance is made for the contribution from heavy nuclei, especially Iron) but these shocks are unable to tap the full power of the explosion because the ejecta have only interacted with a small amount of circumstellar matter.  As the shocks interact with more and more material they slow down and the maximum attainable particle energy drops, but the total acceleration power increases.  It is tempting to identify the `knee'' region with acceleration at the time of sweep-up when the shocks are at maximum power and there is still significant field amplification.  As the shocks continue to expand and slow down the maximum cut-off energy drops as the field amplification becomes less and less effective.  Finally at very late times the shocks weaken to the point where they can no longer maintain the scattering needed for shock acceleration and all the particles inside the remnant begin to diffuse out into the general ISM.  On this picture the steepening of the all-particle energy spectrum at the `knee' is due to the relative lack of power in the very fast early shocks responsible for the highest energy particles combined with the decrease in abundance as one moves to heavier elements.  For the particles below the `knee' we are dealing with shocks that have come into equilibrium and where the entire explosion energy is available.  It is important that, as shown in  \cite{Caprioli2010160, 2011MNRAS.415.1807D}, to first order it is then immaterial when exactly the particles escape.  Thus the cut-off energy can continue to drop without significant impact on the time-integrated production spectrum which remains close to the equi-distribution one with amplitude fixed by the total explosion energy input (although of course with better charge-resolved statistics one might hope to see some spectral features below the `knee', as perhaps hinted at in recent data).

On composition this implies that at the higher energies, that is the `knee' and above, we should be seeing particles accelerated from the immediate surroundings of the supernova progenitor, thus from a highly ionized region with significant contamination by progenitor winds etc.  At the `knee' this should gradually be replaced by a composition more typical of the general ISM and at low energies (where we have the best data!) the composition should be dominated by particles accelerated just before the remnant died and the weakened shock was running into a rather normal and undisturbed ISM.  If one assumes that this is a conventional dusty ISM (as indicated by a broad range of astronomical observations) and allows for modest acceleration and subsequent sputtering of small dust grains it is possible to sketch a quantitative physical model which appears to explain all the features of the chemical composition in terms of standard shock acceleration applied to a dusty ISM; see discussion in 
\cite{2000NuPhA.663..843O, 1997ApJ...487..197E}.

Thus to return to the three-fold origin, on this picture the particles come from the circumstellar and interstellar medium, the energy from the supernova explosion, and the actual acceleration site is the strong collisionless shock driven by the shock and running into the surroundings.  In fact any strong shock in the ISM should contribute to cosmic ray production, not just the forward shocks in supernova remnants, but these are likely to be energetically the most dominant.  In particular supernova remnant models also have reverse shocks running back into the stellar ejecta, and these could in theory contribute material with very interesting compositional signatures, but the power of the reverse shock is much less than that of the forward shock except for a very brief period around sweep-up.   More plausible is that multiple interacting supernova blast waves in super-bubbles make a contribution and colliding strong stellar winds in young OB associations may also be minor sources.

This appears to offer a consistent and plausible account of the origin of Galactic cosmic rays at least as far as the `ankle' which is consistent with the observations and with our general astronomical understanding of the Galaxy.

\section{The ultra high-energy (UHE) particles}

It seems clear that the particles above the `ankle' must have an extra-galactic origin for the reasons discussed above.  Our knowledge of the extreme end of the energy spectrum has been greatly improved in recent years by data coming from the Pierre Auger observatory in Argentina.  Interpretation of these data is still somewhat fluid and subject to revision, however it appears certain that Auger sees a cut-off in the all-particle energy spectrum which could either be the long-expected GZK effect (if the particles are mainly protons) or reflect an intrinsic limitation in the accelerators combined with nuclear photo-disintegration (if the composition is dominated by heavy ions, as the measurements currently seem to suggest).  There is certainly no evidence for significant fluxes of particles at energies beyond the GZK limit, and no evidence for exotic top-down models where these UHE particles result from the decay of primordial ultra-heavy X-particles (corroborated by the lack of evidence for gamma-rays at these energies).  The Auger experiment's report of a significant correlation between the arrival directions of the UHEs and the local distribution of matter still stands, but the correlation is not as strong as initially reported and certainly does not point to any specific class of accelerating system (as yet; with better statistics this may improve somewhat, but almost certainly a definite association will have to wait for a next-generation experiment with substantially greater collecting power than Auger such as the proposed JEM-EUSO \cite{1367-2630-11-6-065009}).  In addition of course there may still be significant deflection by the poorly determined intra-galactic magnetic fields, especially if the particles are heavy nuclei, so that the lack of a strong correlation is not surprising, however it is surely significant that we start to see an anisotropy just at the energy where either the GZK photo-pion losses or photo-disintegration of heavy nuclei restrict potential sources to being relatively nearby in cosmic terms \cite{2011arXiv1107.2055T}.
For an extensive recent discussion of the general issues in the context of one specific model and with many references see \cite{2006PhRvD..74d3005B}.
Although now somewhat dated a good overview and very readable introduction is \cite{2000PhST...85..191B}.

It appears entirely plausible that essentially the same process of diffusive shock acceleration with self-amplified magnetic fields, but operating in the strong jets and outflows known to be associated with active galactic nuclei, can produce these particles \cite{2010MNRAS.405.2810H}.  In addition to diffusive shock acceleration other mechanisms that have been proposed include shear acceleration at the boundaries of the radio jets \cite{2008arXiv0801.1339O}, the `converter' mechanism \cite{2008IJMPD..17.1839D} and stochastic acceleration by turbulence in the radium lobes \cite{2009MNRAS.400..248O}.

Other possible extra-galactic acceleration sites that have been discussed in the literature include gamma-ray bursts \cite{2005ApJ...627..868G} and cluster accretion shocks.  While gamma-ray bursts have enough energy to power the UHE acceleration process \cite{2011arXiv1108.1551L}, it is unclear whether these ultra-relativistic systems would be able to accelerate ions to the required energies, especially in the presence of very strong radiation fields \cite{2010PhyU...53..691P}.
One solution would be to accelerate the UHE particles in a late phase of the evolution after the initial bright flash has decayed away and the shocks have decelerated to being mildly relativistic (which offers some advantages; acceleration at highly relativistic shocks is not that easy \cite{2009MNRAS.393..587P}).  Another interesting idea is to consider semi-relativistic hypernovae (an intermediate class between standard supernova and GRBs) as possible sources \cite{2008AIPC.1000..459W}.

Cluster accretion shocks are the largest shocks in the known universe, but are not particularly strong \cite{2008ApJ...689.1063S} and the magnetic fields are of course very uncertain; nevertheless because of their large length and time scales, and the power available, they are possible sites for the acceleration of UHE particles. For a good recent discussion see \cite{2009arXiv0910.5715V} and references therein.

The bottom line is that with the reduction in the maximum rigidities implied by the Auger results (which have effectively ruled out earlier claims of significant fluxes beyond the GZK limit and also hint at a heavy composition) and the introduction of magnetic field amplification as a key part of shock acceleration there is no shortage of possible extra-galactic acceleration sites.  Indeed it is quite probable that, as in the Galactic case, there are multiple classes of sources and that almost all sufficiently strong collisionless shocks, wherever they occur, can contribute.

\section{Observational tests}

A consistent theoretical picture that `saves the phenomena' is all very well, but ideally as a scientific theory one would like to be able to make specific predictions that are capable of observational verification (or, perhaps more importantly, falsification).  Several years ago at an ISSI workshop a group of us attempted to compile such a list \cite{2001astro.ph..6046O}
and it is interesting to revisit this list and see what progress has been made.  On rereading this article some things are immediately apparent.  Firstly, it was written before the idea of magnetic field amplification in shocks became widely accepted, which solves one of the major problems it identifies (the particles between the `knee' and the `ankle').  Secondly, it preceded the definitive detection of TeV gamma-ray emission from a number of shell-type SNRs which has conclusively demonstrated that at least some young SNRs are accelerating charged particles to energies of order $10^{14}\,\rm eV$.  It remains unclear whether these are mainly protons or electrons, however recent observations connecting TeV and GeV gamma-ray emission to molecular clouds near SNRs \cite{2011arXiv1104.1197U} are a strong hint that hadronic processes are involved and thus that ions are being accelerated. It remains worrying however that there is no unambiguous detection of a Galactic Pevatron in high-energy gamma-rays although this can perhaps be explained if the Pevatron phase is relatively short-lived.

Nevertheless it is interesting that the best evidence available then \cite{2000ApJ...543L..61H} for efficient cosmic ray production in SNRs was essentially the same as now; that there is `missing energy' in some young SNRs in the sense that the thermal energy deduced from the X-ray temperatures is much lower than that expected from proper-motion estimates of the shock velocity and the only plausible reservoir for this `missing energy' is a cosmic-ray component produced in the shock
\cite{2009Sci...325..719H}.  The multi-wavelength modeling of certain remnants is also beginning to yield convincing arguments in favor of strong cosmic ray acceleration, e.g. \cite{2010A&A...511A..34B,2011arXiv1105.6342M}, although in other cases the evidence is ambiguous or favours an electron dominated scenario \cite{2011ApJ...735..120Y,2011ApJ...734...28A} .  The decisive proof would of course be the detection of high-energy neutrinos associated with either a SNR or (more likely) a SNR interacting with molecular clouds.
This is challenging, but may be possible soon with, e.g., the proposed KM3NeT observatory.

As to the ultra-high energy extragalactic particles, it is probable that we will have to wait for a next-generation experiment to definitively pin down their origin.

In conclusion the author would like to thank the editor, Pasquale Blasi, for his patience and for helpful comments which, together with those of the referees, have materially improved this article.

\onecolumn

%\bibliographystyle{abbrv}
%\bibliography{review}

\printbibliography

\end{document}